Research Article

# Some Old Globular Clusters (and Stars) Inferring That the Universe Is Older Than Commonly Accepted

Félix Llorente de Andrés[1, 2, *]

[1]Department of Astrophysics, Center of Astrobiology (CAB) - European Space Astronomy Center (ESAC) Campus, Villanueva de la Cañada (Madrid), Spain

[2]Section of Science and Technology, Almagro Athenaeum, Almagro (Ciudad Real), Spain

## Abstract

The James Webb Space Telescope (JWST) has made startling discoveries regarding the early universe. It has revealed galaxies as soon as 300 million years after the Big Bang, challenging current galaxy formation models. Additionally, it has identified massive, bright galaxies in the young universe, contradicting the standard ΛCDM model's age estimate of 13.8 Gyr. This prompts a re-evaluation of galaxy formation and cosmological models. There is a strong tension between JWST high-redshift galaxy observations and Planck Cosmic Microwave Background (CMB) satellite measurements. Even alternative cosmological models, including those incorporating dark matter–baryon interaction, f(R) gravity, and dynamical dark have failed to resolve this tension. One possible solution is that the Universe's age exceeds predictions by the ΛCDM model. The study challenges this by introducing a method based on blue straggler stars (BSs) within GCs, comparing ages with other models. The ages obtained are compared with those of other models to certify that they are equally valid. These values are comparable within the error ranges except for the clusters: NGC104, NGC 5634, IC 4499, NGC 6273 and NGC 4833, finding their respective ages to be between 14.7 and 21.6 Gyr, surpassing the commonly accepted age of the Universe. These results inferred an age for the Universe of around 26 Gyr, close to 26.7 Gyr. This value aligns that suggested by the cosmological model named Covarying Coupling Constants + TL (CCC+TL). Such a value is consistent with early universe observations from the James Webb Space Telescope (JWST). The results of the present paper reinforces the advocating for a critical review of models encompassing dark mass, dark energy, and the dynamics of the Universe, particularly in explaining the presence of primitive massive galaxies, very old GCs, and very old and poor metallic stars.

## 1. Introduction

The determination of the age of the oldest clusters is so important insomuch as of these ages becoming the most stringent lower limits on the age of our galaxy, and the Universe as well. However stellar models and the methods for the age determinations of globular clusters are still in need of improvement.

There are three independent ways to reliably infer the age of the oldest stars in our galaxy: nucleocosmochrology, white

---

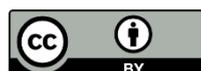

dwarf cooling curves, and the main sequence turnoff time scale. The first way is based on the Thorium (232Th) and Uranium (238U) abundance measurements and a comparison of these abundances to the abundance of other r-process elements. This method presents difficulties because Th and U have weak spectral lines so this can only be done with stars with enhanced Th and U (requires large surveys for metal-poor stars) and unknown theoretical predictions for the production of r-process (rapid neutron capture) elements. Based on this method, an age of 13.84 ±4 Gyr is found for the star metal-poor star BD +17 3248 [1]. For CS 31082-001, the Uranium abundance alone gives an age of 12.5 ±3 Gyr while the Th/Ur ratio of this same star gives 14.0 ± 2.4 Gyr [2]. These ages are consistent with those obtained by other methods. The second method, white dwarf cooling curves, was performed by Hense et al (2007) [3] based on a deeper exposure of NGC 6397 with HST (Hubble Space Telescope). They estimate age for this cluster of 12.7 ±0.7 Gyr. This age is compatible with ages found for other globular clusters by means of the third method: the main sequence turn-off. This method the method is based on the luminosity at this "main sequence turnoff" point and exhibits the smallest theoretical uncertainties.

However it has been found that the uncertainty in translating magnitudes to luminosities (i.e., the uncertainties in deriving distances to globular clusters) is the main source of uncertainty in globular cluster age estimates. Some examples would be: Gratton et al (2003) [4] find for NGC 6397 & NGC 6752 ages of 13.9 ±1.1 and 13.8 ± 1.1 Gyr respectively. For 47 Tuc (NGC 104) the age would be 11.3 ±1.1 Gyr and 13.4 ± 0.8 Gyr for the oldest clusters.

Nevertheless, there are another methods as the recent study performed by Florentino et al (2016) [5] combining the Cosmic Microwave Background and Baryonic Acoustic Oscillations and the method of turn off (TO) found age values for NGC 2808 of 10.9 ±7 Gyr. Bono et al (2010) [6], based on the difference in magnitude between the main-sequence turnoff (MSTO) and the lower main sequence (MSK), estimate for NGC 3201 and age estimates 11.48 ±1.27 and 11.55 ± 1.53 Gyr. Massari et al (2016) [7] using deep IR found that NGC 2808 has an age of 10.9 ± 0.7 (intrinsic) ±0.45 (metallicity term). And just for finalizing the frame, Correnti et al (2016) [8] from the Hubble Space Telescope Wide Field Camera 3 IR archival observations of four GCs—47 Tuc (NGC 104), M4 (NGC 6121), NGC 2808, and NGC 6752—for which they derived the fiducial lines for each cluster and compared them with a grid of isochrones over a large range of parameter space, allowing age, metallicity, distance, and reddening to vary within reasonable selected ranges. The derived ages for the four clusters are, respectively, 11.6, 11.5, 11.2, and 12.1 Gyr.

If all these results are used to constrain cosmological parameters, one needs to add to these ages a time that corresponds to the time between the Big Bang and the formation of globular clusters in our galaxy. Let it be supposed that the age of the Universe is 13.8 Gyr, margin 20%, which means: the oldest clusters should be younger than 13.6 Gyr; this figure should be became the upper limit. Interestingly, for the best fit value of the Hubble constant, globular cluster age limits also put strong limits on the total matter density of the Universe; but it is no matter of this work.

Very recently, JWST observations have unveiled a surprising revelation regarding the early formation and evolution of stars and galaxies, occurring much earlier than initially anticipated (see references in Gupta (2023) [9]. These observations not only confirmed the existence of galaxies as early as 300 Myr after the Big Bang but also revealed a higher number density than predicted by current galaxy formation models. Furthermore, these findings reported the existence of massive, bright galaxies in the very young Universe, with an estimated age of approximately 13.8 Gyr according to the standard ΛCDM model. These discoveries raise a fundamental question: How did galaxies in the very early Universe evolve to a degree comparable to those with billions of years of evolution, some as early as less than approximately 300 Myr after the Big Bang? This question necessitates an exploration of adjustments to well-established galaxy formation and cosmological models, originally developed to explain observations at lower redshifts. Wang and Liu (2023) [10] noted a strong tension between JWST high-redshift galaxy observations and Planck CMB measurements. Even alternative cosmological models, including those incorporating dark matter–baryon interaction, f(R) gravity, and dynamical dark energy (Santini et al. 2023) [11], have failed to resolve this tension. Steinhardt et al. (2023) [12] demonstrated the presence of stronger Balmer breaks extending beyond z > ~11, challenging the validity of early galaxy formation templates at high redshift and suggesting the need for new physics beyond the ΛCDM model. Gupta (2023) [9] through a series of studies, proposed an extension of the ΛCDM model named CCC + TL. This model, hybridized with the tired light concept and parameterized with Pantheon + data, aligns with deep space observations from JWST. Gupta's assertion is that the age of the Universe is 26.7 Gyr, with 5.8 Gyr at z = 10 and 3.5 Gyr at z = 20, allowing for the formation of massive galaxies.

The age proposed by Gupta (2023) [9] for the Universe eliminates the need for lower limits on the ages of clusters and the oldest stars. Debates regarding the proposed older age of the Universe, often based on the ages of globular clusters and oldest stars, become model-dependent as the age of globular clusters adjusts with evolving models. For instance, Bolte and Hogan (1995) [13] determined certain cluster ages to be 15.8±2.1 Gyr, while Tang and Joyce (2021) [14] changed, based on MESA stellar evolution code, its best age estimate, 14.5±0.8 Gyr, to a lower 12.0±0.5 Gyr. More recently, Plotnikova et al 2022 [15] based on standard isochrones fitting they found that more than half of their old star sample exceeding the age of 13.8 Gyr (see Figures 12 and 15 of [16]).

The lack of a universal age scale spanning the entire time domain, coupled with the absence of systematic and robust cross-calibration among the various methods employed, poses a challenge in constraining stellar ages below 13.8 Gyr. Jeffries et al. (2023) [16], unburdened by age constraints on recently born clusters, adjusted the age of the young open cluster IC 4665 from 32 million years to greater than 50 million years. Hence, the age of a star or cluster cannot be considered a constraint on the age of the Universe. The primary impediment lies in the absence of a universal age scale and a systematic cross-calibration of contemporary methods. Gupta's model extended age eliminates the need for lower limits on the ages of globular clusters and oldest stars, challenging the conventional criticism based on their ages (Cimatti and Moresco 2023) [17].

This work attempts to obtain a more objective method of age determination based on the abundance of blue straggler stars (BSs). This way of deriving globular cluster ages is attempted for the first time. It is not an irrational idea because derived ages based on the turn-off are worthy ages. Why not a method based on the number of BSs wouldn't work. This work tries to obtain new method of age determinations. This method has, as the others do, some inconsistencies and is not trouble free. Anyway, the method contributes to obtain very consistent ages. Given the lack of a universal age reference, the robustness of the model is demonstrated by comparing the most recent results published by other authors.

The objective of this study is to identify, from a set of globular clusters, those with ages surpassing the currently accepted age of the Universe. The ages of these globular clusters are deduced by applying the model developed by Llorente de Andrés and Morales Durán (2022) [18], but be ensured that the model is applicable to GCs. As described in Section 2 This Section is devoted to the description of the model and the used methodology.

The validation of the model is tackled in the Section 3, while the results are debated in Section 4. Finally the conclusions are exposed in Section 5.

## 2. Description of Models and Methodology

Discrepancies between the ages of the oldest stars and the cosmologically inferred age of the universe are well-documented and may be influenced by systematic effects in the predicted evolution of low-mass stars (Valls-Gabaud, 2014) [19]. To address this, recent models aim to mitigate these effects and reduce uncertainty in the age determination of the Universe derived from GCs and oldest stars, thereby avoiding constraints on cosmological models.

This section is devoted to describe the model used in this work to deduce the age of the sample of GCs, labeled as this work age in Appendix Table A1, and their ages independently of the stellar evolution model. In advance it can also be affirmed that the metallicity does not play any relevant role. The [Fe/H] values have been added to Appendix Table A1. These values are sourced from Bailin (2022) [20].

### 2.1. Description of Models

In a previous work (Llorente de Andrés and Morales, 2022) [18], called hereinafter LM, it was explained that the presence of BSs in a cluster is not accidental, nor random, but its discrete number follows a function which is related to the ratio between age of the cluster and relax time (called $f$ = age/trlx) and to a factor, we call $\varpi$, which is the indicator of stellar collisions plus primordial binary evolution thus related to the total number of observed BSs (NBS). The basis of this model is the solution of the dynamic equation of the star cluster: part expansion of the cluster and part concentration or core collapse. The line of reasoning is made explicit through Figures 7, 8 and 9 published in [18]. Thus the first step is to verify that this model is equally applicable to globular clusters. For corroborating results of presence of BSs, it was applied the function (equation (1) in the present paper), corresponding to the equation 6 of the referenced paper [18], to the complete set of observed number of BSs of globular clusters (GC) listed by Moretti et al (2008) [21], hereinafter MAP.

$$NBS \cong f^3 \left(\frac{1}{e^{f/\varpi}-1}\right) \quad (1)$$

The selection of MAP is due to their catalogue of BSs extracted from a homogeneous sample of 56 Galactic globular clusters (GC) observed with Wide-Field Planetary Camera 2 on board of Hubble Space Telescope (WFPC2/HST). MAP established relations between the frequency of BSs and the main parameters of their host GC, concluding that any population of BSS strongly depends on the luminosity of the cluster, on the extension of the cluster horizontal branch, and on the central velocity dispersion. Moreover, they find that clusters having higher mass, higher central densities, and smaller core relaxation timescales have, on average, more BSs; while BSS in fainter clusters are mostly influenced by the cluster luminosity and the dynamical timescales. The GC sample proposed by MAP is suitable for testing the model, represented by equation (1), hereinafter named LM model, as the study aligns with the advancement of such model.

Based on equation (1), the number of theoretical BSs is calculated by assuming that the relax time (trlx) might be the value alike to that one listed as the median relaxation time trlx published by Harris (1996; but edition 2010) [22], herein after H10. To avoid bias because of a single source, the present study was also performed by taking the median relaxation time from Recio-Blanco et al (2006) [23], herein after RB. As GC median age it was adopted the value suggested by Kraus and Chaboyer (2003) [24], herein after KC, 12.6 Gyr.

In order to extend the data, not restricted to only one catalogue, this work was also carried out with another BSs catalogue published by Leigh et al (2007) [25], herein after LSK. They use a large, homogeneous sample of BS stars of 57 globular clusters to investigate the relationships between blue straggler populations and their environments. Their results are inconsistent with some models of BSs formation that include collisional formation mechanisms suggesting that almost all observed blue stragglers are formed in binary systems.

By comparing LM theoretical results of the number of BS stars with two methods of star detection, thus the positive outcome of this comparison strengthens the robustness of the model; in spite of the difference in the number of BSs between the two catalogues. This difference comes from the different normalization applied in each identification and collection of BSs. For instance, MAP normalize absolute number of BSs in a region divided by the total luminosity of stars in the same region. However, LSK normalize the BSs absolute number to different cluster populations. This difference will be very useful for establishing a minimum and a maximum for the GC ages.

Thus, it was computed, cluster by cluster, by means of the above mentioned equation (1) the number of predicted BSs. By making use of the two median relaxation times (relaxation time is the scale of time where the star changes its track) taken from H10 and RB respectively and an adopted median age for GCs of 12.6 Gyr, as suggested by Kraus and Chaboyer (2003) [24].

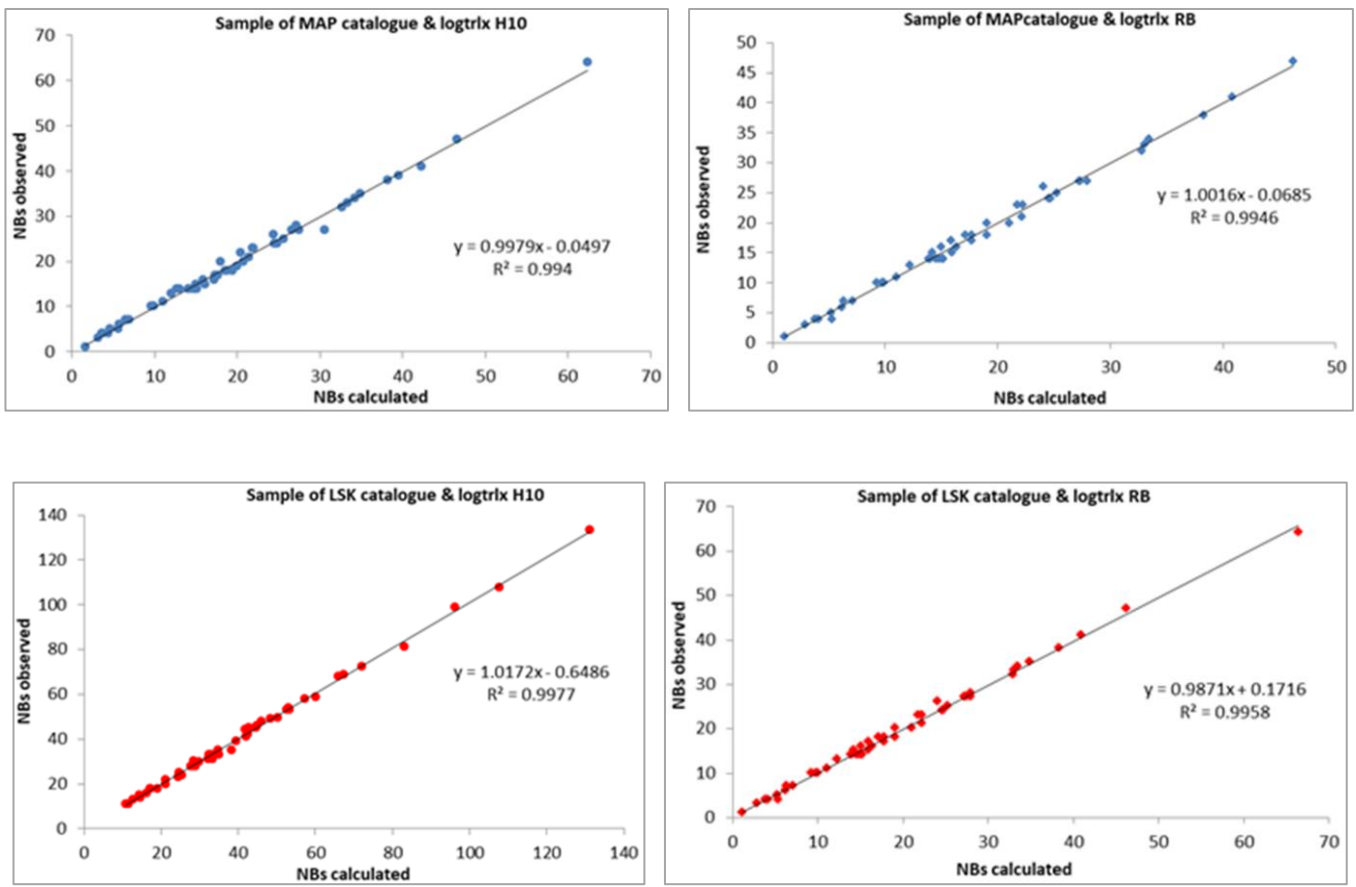

*Figure 1.* This graphics showing that the number of BS stars computed from the theoretical/empirical LM's equation 6 of [18] match with number of observed number of BS stars. Top: NBs calculated from MAP catalogue and H10 trlx. Middle top: NBs calculated from MAP catalogue and RB trlx. Midlle bottom: NBs calculated from LSK catalogue and H10 trlx. Bottom: NBs calculated from LSK catalogue and RB trlx.

These results are displayed in Figure 1 This represents the number of predicted BSs versus the number of observed BSs for both MAP and LSK catalogues. In all the cases, the relation was found one to one with an $R^2 \cong 1$, in whatever the catalog.

The conclusion is that the distribution function holds, allowing the deduction of cluster ages. Previously, to a certain extent, Ferraro et al. (2020) [26] already made an approximation to the relationship between the number of BSs and the age of the CG through the relaxation time, reminiscent of the reasoning presented by LM in their section devoted to the definition of f, where they explained how BS stars are formed. Firstly, a definition of a ratio that might be equivalent to the 'frequency or probability of encounters and/or deflection of orbits' based upon the crossing time. Thus, crossing time and relaxation time are related to the number of BS stars. f is the ratio between relaxation time and the age of the GC, indeed an indicator of the probability of encounter and/or deflection of orbits. This ratio f is implicitly normalized to the total number of stars, N; thus, f can be compared among clusters, irrespective of their mass.

## 2.2. Methodology

Let's summarize, the equation (1) establishes a relationship between the number of BSs in a GC with the factor $\varpi$, and with the trlx and age – the latter two variables through the factor $f$. It should be straightforward to determine the corresponding age for the GC sample. This determination is made using the relationships between $[f]$, $[\varpi]$ – specifically, the median of $f$ and $\varpi$. Figure 2 illustrates this relationship for the four potential cases: MAP and LSK catalogues, and trlx H10 and RB. It's evident that the relationship between $\varpi$ and $f$ is accurately represented by a linear function (except for when $f$ is less than ~4-6 and $\varpi$ is greater than ~3.5-4 - in which case, a distinct approach is required). For each GC, the relationship with the best fit equation is selected, resulting in a new value of $f$, denoted as $f_T$. It's this $f_T$ value that facilitates the derivation of the GC age (age = $f_T$ * trlx). The process initiates with an exceptional stage: the portion of the graph displaying significant dispersion (refer to Figure 2). These values are set aside, and the process continues with the remaining data points.

At this point it would be necessary to consider a possible cut off because of it is assuming that the globular clusters are not older than 13.6 Gyr which would be compatible with an age for the Universe of 13.8 Gyr. However, the process is not stopped in those clusters whose computed ages are higher than 13.6 Gyr. The present study's GC sample is treated completely. The process is repeated, with all the values. At each step, a new equation is solved then a new $f_T$ is obtained yielding to corresponding value for age. In consequence some GCs are older than the common adopted age for the Universe.

## 3. Validation of the Model

This section is devoted to the validity of deduced ages of the present study by comparing with two sets of globular clusters (GCs) from different samples. The model's robustness is demonstrated by comparisons with two most recent works.

Comparisons with other recent works, such as Valcin et al. (2021) [27] (hereinafter VA) and Usher et al. (2019) [28] (hereinafter US), highlight the robustness of model above described. The lack of a universal age scale and the absence of systematic cross-calibration methods contribute to uncertainty in GC ages. [27] reduces uncertainty in age determination using precise metallicity determinations, inferring a Universe age of ~13.5+0.16-0.14 Gyr (at 68% confidence level), in agreement with Planck mission's model-dependent value of 13.8 ± 0.02 Gyr. [28] deduced GC ages from MIST isochrones, developed by [29] Choi et al. (2016) [29]. These age values are also below 13.5Gyr. The three studied GCs and the corresponding ages are displayed in Appendix Table A1.

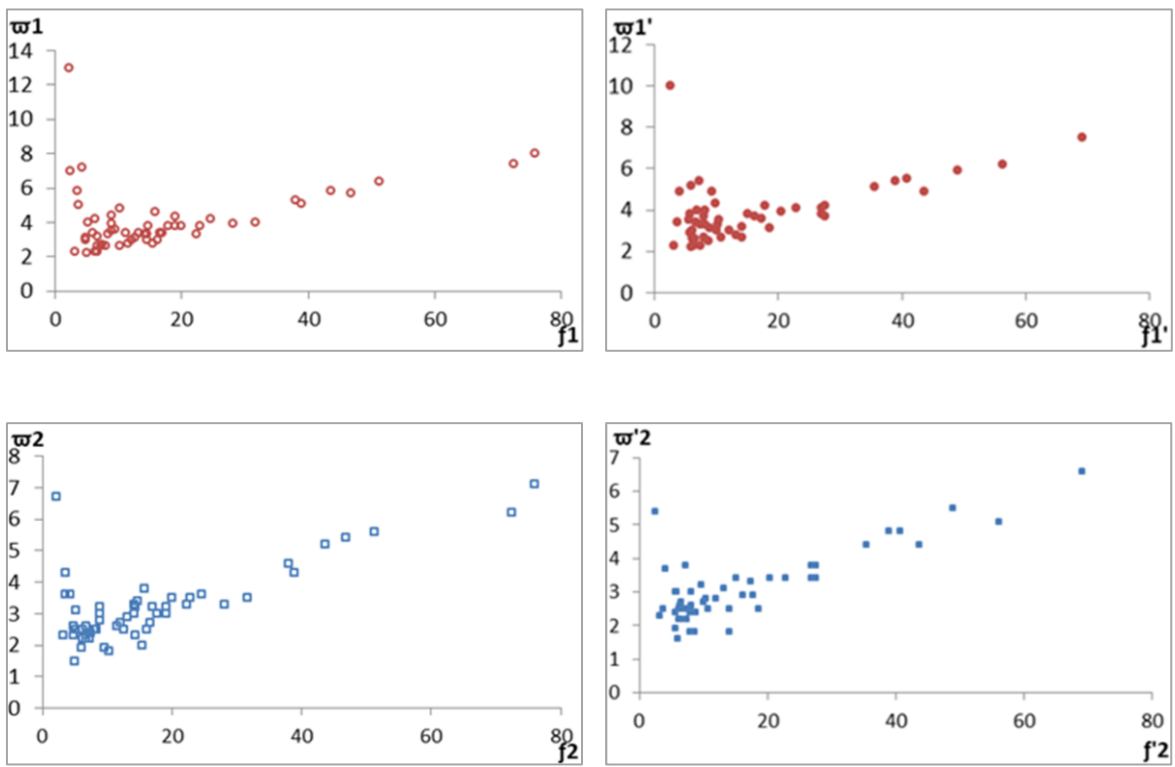

*Figure 2.* This figure shows the relationships between $\varpi$ and $f$ for the four cases. Top: factors derived from MAP catalogue and H10 trlx. Middle top: factors derived from MAP catalogue and RB trlx. Midlle bottom: factors derived from LSK catalogue and H10 trlx. Bottom: factors derived from LSK catalogue and RB trlx. In the four graphics it is easy to distinguish two separate regions; $f$ < 5-6 and $\varpi$ > 4 and 3; the difference in the last value is due to the different number of BSs listed in both catalogues.

The age dependence on [Fe/H] is displayed in Figure 3. This Figure 3 illustrates that the model of the present study, unlike others, does not show a dependence on metallicity.

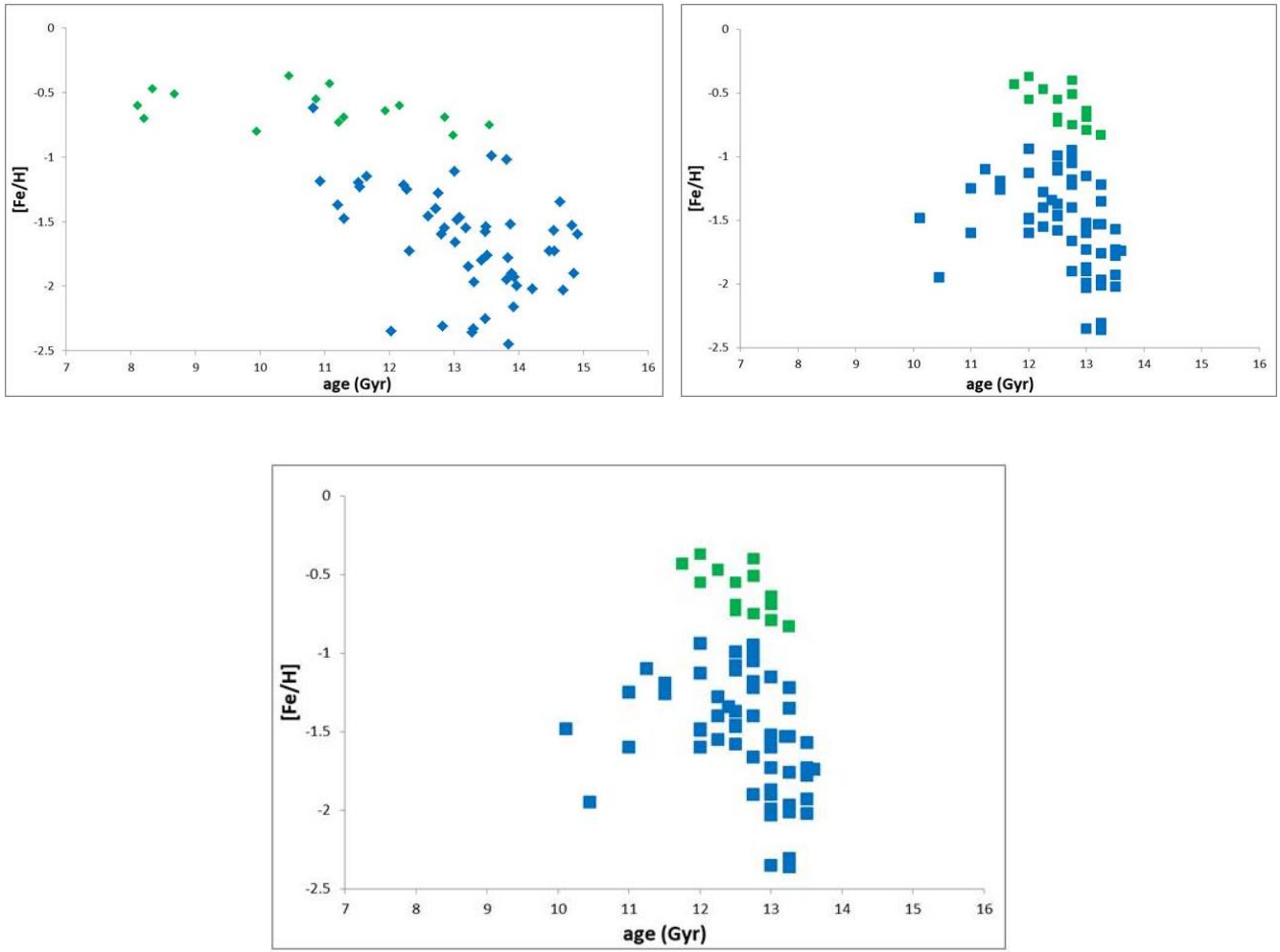

*Figure 3. This figure represents the relationship between metallicity and age, as deduced by each author. [25] (diamonds) show a dependent relationship, while [26] (squares) and present study model (dots) display independence from metallicity. Green symbols highlight the gap at [Fe/H] ∼-1.0 dex. That is less evident in this work model.*

In summary, the model's robustness has to be demonstrated by comparisons between the GC ages from this model and those obtained from the most recent works and its independence from systematic effects in low-mass star evolution sets it apart from other age-determination models. It assesses the validity of deduced ages by comparing the three sets of globular clusters (GCs) from different samples: VA with 67 GCs, US with 75, and this work with 49. Commonalities between samples include 28 GCs with VA, 46 with US and 53 between VA and US, providing a robust basis for comparison.

It is tantalizing to compare age values computed by different authors who made them from different, and sometimes, no related methods. We use a statistical method, called the Bland-Altman diagram, which is a powerful tool to evaluate the agreement between measurement techniques and therefore their results. These plots illustrate the difference between two measurements on the y-axis and the average of the two measurements on the x-axis.

In the present study, the Bland-Altman plots corresponding to the ages compared between the three methods, this work, VA's and US's, are depicted in Figure 4, focusing on GCs common to each method. The plot includes lines representing ±2σ standard deviations of the differences, aiding in identifying outliers. Moreover lines representing 1σ is including. Bland-Altman plots are instrumental in detecting systematic bias or random errors in the data. If the scatter of points surpasses the twice standard deviation, it may suggest the presence of random errors. Additionally, outliers falling outside the twice standard deviation lines can be identified. Only NGC 5986, NGC 6362 and NGC 6681 appear to be out of the 2σ frontier.

The application of the Bland-Altman plot to the present study distributions supports the conclusion that all three methods are equally valid for deriving the ages of their respective GCs. So, Figure 4 shows that the inferred ages ex-hibit considerable similarity despite stemming from those obtained by [27] and those obtained by [28]. This allows to propose that the ages derived from the present study are equal confidence as those of the other authors.

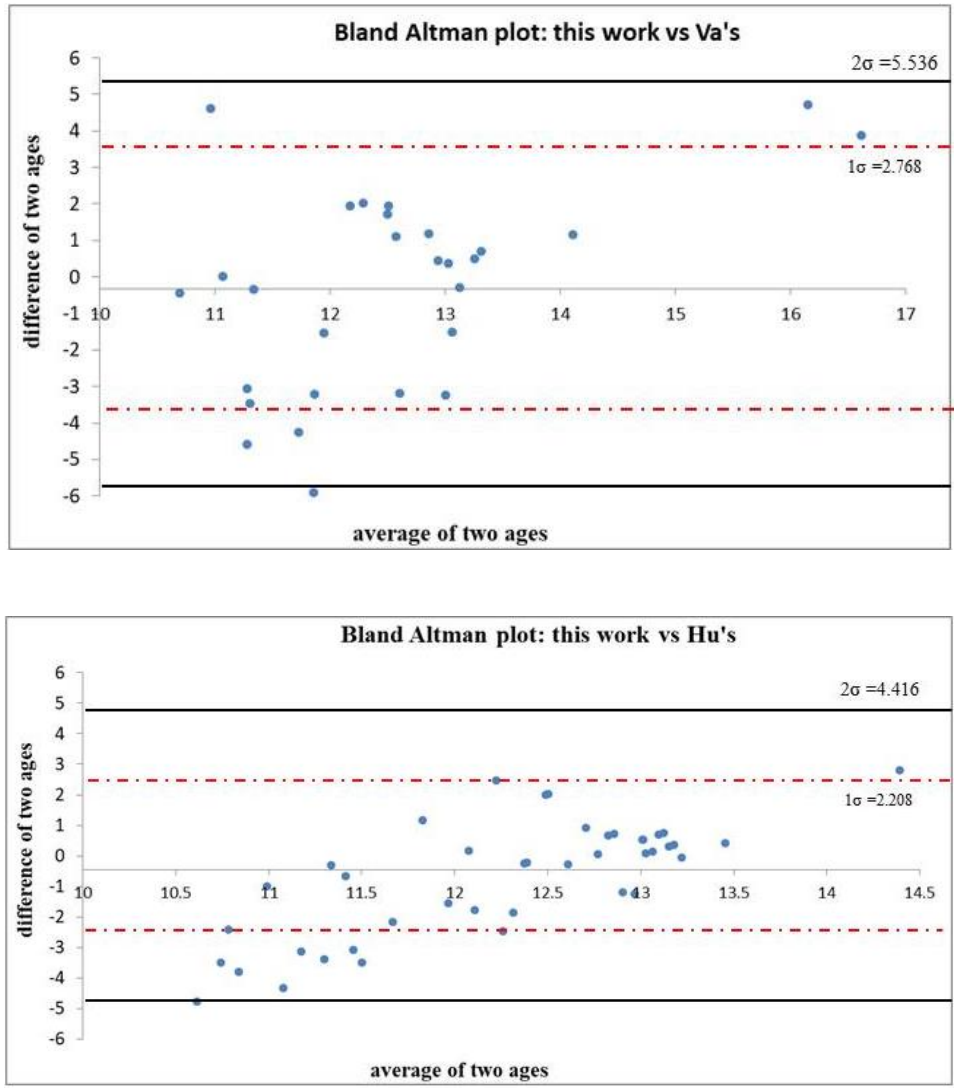

***Figure 4.*** *This figure shows the Bland-Altman plot that shows that the age distributions are comparable, both this work compared to that of [25] and compared to that of [26]. The origin of the axis corresponds to the average of the mean of the values of age for both methods, mentioned in the title. The black straight line represents the 2σ separation and the red dashed line to the 1σ separation. This figure made for those GCs that are the same to each method.*

Moreover an illustration of the confidence, in addition to the conclusions drawn from the Bland-Altman plot, is evident in the comparison of the mean values from the three methods, with LM method yielding the lowest values. Thus, for the average age across the entire sample, [27] (Valcin's) is 12.7 Gyr, [28] (Usher's) is 12.5 Gyr, and the present work is 12.1 Gyr. The difference in mean ages among the three is smaller than the errors taken from [27] (the best-calculated/estimated ones), ranging between +0 and +2.76 Gyr and between -0.45 and -1.80 Gyr. If considering only the clusters common to the sample of the actual paper and the other authors, the mean ages are practically the same: for this actual sample of clusters common with [27] is 12.4 vs 12.8Gyr, for actual paper clusters common with [28] is 12.2 vs 12.7Gyr and for [27] vs [28] their medians are, respectively, 12.7 and 12.3Gyr. The difference between the medians is smaller than the smallest of the sigma's, σ ~0.625, calculated in the common clusters.

In conclusion, the three methods are equally reliable.

*Table 1. Calculated ages, from this work, for assumed GCs older than 13.8Gyr.*

MAP catalogue

$f1_T = 7.08 - \varpi1/1.98$; $R^2 = 0.52$

| Cluster | Logtrlx H10 | $f1$ | $\varpi1$ | NBS Cal | NBS Obs | $f1_T$ | age1$_T$ |
|---|---|---|---|---|---|---|---|
| IC 4499 | 9.73 | 2.35 | 7 | 32 | 33 | 3.54 | 19.04 |
| NGC 104 | 9.55 | 3.55 | 5.8 | 53 | 54 | 4.15 | 14.73 |
| NGC 5634 | 9.54 | 3.63 | 5 | 45 | 46 | 4.55 | 15.79 |

LSK catalogue

$f1_T = 11.65 - \varpi1/0.9384$; $R^2 = 0.62$

| Cluster | Logtrlx RB | $f1$ | $\varpi1$ | NBS Cal | NBS Obs | $f1_T$ | age1$_T$ |
|---|---|---|---|---|---|---|---|
| NGC 104 | 9.48 | 4.17 | 4.9 | 54 | 54 | 6.43 | 19.41 |

LSK catalogue

$f2_T = 16.961 - \varpi2/0.3354$; $R^2 = 0.48$

| Cluster | lLogtrlx H10 | $f2$ | $\varpi2$ | NBS Cal | NBS Obs | $f2_T$ | age2$_T$ |
|---|---|---|---|---|---|---|---|
| NGC 104 | 9.55 | 3.55 | 4.3 | 35 | 35 | 4.14 | 14.70 |
| NGC 5634 | 9.54 | 3.63 | 3.6 | 28 | 27 | 6.23 | 21.6 |
| NGC 4833 | 9.42 | 4.79 | 2.6 | 20.69 | 20 | 7.05 | 18.6 |
| NGC 6273 | 9.38 | 5.25 | 3.1 | 33 | 32 | 7.72 | 18.5 |

LSK catalogue

$f2_T = 19.401 - \varpi2/0.2918$; $R^2 = 0.53$

| Cluster | Logtrlx RB | $f2$ | $\varpi2$ | NBS Cal | NBS Obs | $f2_T$ | age2$_T$ |
|---|---|---|---|---|---|---|---|
| NGC 104 | 9.48 | 4.17 | 3.7 | 35 | 35 | 6.72 | 20.3 |

## 4. Results and Discussion

The validation of exposed method allows to resume age calculations, extracting those exceeding the assumed age for the Universe. Table 1 shows cases GC ages exceeding 13.8 Gyr. The Table 1 includes the linear correlation between $\varpi$ and $f$, their corresponding $R^2$ values. The GCs exceeding 13.8 Gyr include NGC 104 (14.7–20.3 Gyr), NGC 5634 (15.79–21.59 Gyr), NGC 4833 (18.55 Gyr), NGC 6273 (18.5 Gyr) and IC 4499 (19.0 Gyr).

The age and metallicity of NGC 104 prompt an unconventional explanation and exotic explanation. Either the existence of an intermediate black hole in the center of the cluster as it is suggested by Kizilton et al (2017) [30]. Or more recently Ke Qin et al (2023, 2024) [31, 32] simulations showed that black hole main-sequence star (BH-MS) binaries with an initial orbital period less than the bifurcation period can evolve into ultra-compact X-ray binaries (UCXBs) that can be detected by LISA challenging the current assumption that its age suggests the Universe is older. Or the age of NGC 104 suggest that the Universe is older than 13.8Gyr. However, the age deduced for GC NGC 5634 leans towards supporting the argument that the Universe is older than commonly accepted.

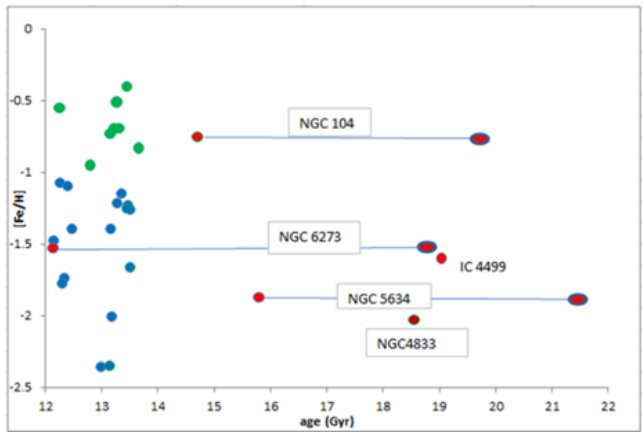

*Figure 5.* *This figure, [Fe/H] vs this work GC age, shows the maximum and the minimum values of the four GCs with age values longer than that assumed to be the age of the Universe. Even in the case of NGC 104 and NGC 5634, their respective minimum value is over 13.8Gyr.*

As mentioned earlier, employing two distinct catalogues—MAP, where more BSs are identified for each of the globular clusters (GCs), compared to LSK—facilitates the derivation of maximum and minimum age values for the respective GCs. For those GCs surpassing the assumed age of the universe, the maximum and minimum ages are visually represented in Figure 5. Notably, IC4499 and NGC 4833 provide singular values which exceed that of the other GCs.

In Figure 5, the range between maximum and minimum ages accentuates the variability and uncertainty in age determination for each GC. This approach, utilizing two catalogues with differing BS identifications, adds robustness to this work method for age calculations by acknowledging the inherent variations in data collection methodologies.

The collective data presented by VA yields a universe age of 13.5 Gyr, consistent with the adopted value (13.8Gyr) with a reliability of 68%. However, nine of their clusters surpass the age of 13.8 Gyr but do not exceed 15 Gyr, These nine GCs are marked in bold in Appendix Table A1.

If that is to apply to the sample of this present study taking 21.59 Gyr, as it is the maximum age of clusters and 19.40 Gyr as the average of the maximum values, the inferred age for the Universe is around 26 Gyr, close to 26.7 Gyr, as proposed by [9].

*Table 2.* *Stars whose ages exceeding 13.8 Gyr quoted from Plotnikova et al (2022) [15].*

| Star Identification | Min&Max age Gyr |
|---|---|
| HE0023-4825 | 14.4--14.9 ±1.2 |
| HE0023-4825 | 13.8--16.0 ±2.9 |
| HE 1052-2548 | 15.3--15.7 ±0.7 |
| HE 1225-0515 | 14.6--14.7 ±1.8 |
| HE 2347-1254 | 14.4--14.6 ±1.7 |
| HE 0231-4016 | 13.3--15.4 ±1.3 |
| HE 0926-0508 | 14.8--14.9 ±0.8 |
| HE 1015-0027 | 14.9--15.1 ±1.3 |

Not only is there the possibility that there are Globular Clusters (GC) whose age surpasses the age of the Universe, as showed above, but also stars that are very metal-poor and very old, to the extent that some of them exceed the age of 13.8 Gyr. A group of these stars has been studied by Plotnikova et al (2022) [15], who, based on the isochrone fitting method using Padova and BaSTI isochrones, have found that in their sample, there are stars with ages exceeding 13.8 Gyr; these stars are listed in Table 2. The exposed data is a summary of the ages obtained from Padova, from BaSTI, and from the mean of both, but only the maximum and minimum values of each star are presented, as well as the minimum and maximum values of the errors. The values are expressed in Gyr.

It is imperative to recognize the potential impact of these variations on age determinations, as well for GCs as for the above listed stars, exceeding the assumed age of the universe. That leads to a review of the standard ΔCDM model. That means that these ages put into question the expansion age for a flat Universe. This would be aligned with the model presented by [9] whose model asserts that the age of the Universe is ~26.7 Gyr; much older than considered until now.

## 5. Conclusions

In the present study, we employed the method LM to compute GC ages, disregarding metallicity and age limitations. Their study is based on the dynamic analysis of the cluster leading us to the relation between the number of observed blue straggler stars (BSs) and the age of the parent GC. Bland-Altman plots confirmed the compatibility of their deduced ages with those of other authors, with exceptions for GCs surpassing 13.8Gyr.

The results are detailed in Table 1 and illustrated in Figure 5, indicate two GCs in our sample whose minimum ages exceeding the currently assumed age for the Universe: NGC104 (14.7–20.3 Gyr) and NGC5634 (15.8–21.6 Gyr). The special case of NGC104 invites exploration of unconventional explanations, challenging the assumption that its age could imply an older Universe. Conversely, the age deduced for GC NGC 5634 with a low metallicity of -1.87 aligns with the proposition that the Universe is older than conven-tionally accepted. Taking into consideration the age of IC 4499 (19.0 Gyr). NGC4833 (18.55Gyr) and NGC 6273 (18.5 Gyr) further support this conclusion.

This conclusions is reinforced by the performed study made by [15], who, based on the isochrone fitting method using Padova and BaSTI isochrones, found that, in their sample of very metal-poor and very old stars, some ages vary between 13.3 and 16.0Gyr. Table 2 listed the maximum and minimum values and errors of each star. The values are expressed in Gyr.

The recent observations from the James Webb Space Telescope (JWST), while confirming the existence of massive, bright galaxies in the very young Universe, present a significant challenge to the accepted age of 13.8 Gyr according to the standard ΛCDM model. In this sense Gupta (2023) [9] proposed a hybrid model incorporating tired light theory and a covarying coupling constants' parameter instead of the cosmological constant in the ΛCDM model. This Gupta's model posits an age of 26.7Gyr for the Universe, aligning with results showed in the present study.

The results of the present paper reinforces the advocating for a critical review of models encompassing dark mass, dark energy, and the dynamics of the Universe, particularly in explaining the presence of primitive massive galaxies, very old GCs, and very old and poor stars. The imminent James Webb Space Telescope stands poised for an in-depth review of a comprehensive set of globular clusters and old stars, offering the potential for further insights and re-finements to our understanding of the broader cosmology field. Furthermore, it can be ventured that the CCC+TL model eliminates the need for dark matter. It will have to wait for the information provided by NASA's upcoming Nancy Grace Roman Space Telescope and ESA's Euclid observatory to clarify this statement.

## Abbreviations

BSs: Blue Straggler Stars
CCC: Covarying Coupling Constants
GC: Globular Cluster
HST: Hubble Space Telescope
HSTWFPC: Hubble Space Telescope Wide Field Camera
IR: Infrared
JWST; James Webb Space Telescope
MESA: Modules for Experiments in Stellar Astrophysics
MIST: MESA Isochrones and Stellar Tracks
MSL: Lower Main Sequence
MST: Main Sequence Turn off
TL: Tired Light
TO: Turn off
ΛCDM: Cosmologic Constant and Cool Dark Matter

## Acknowledgments

The author expresses gratitude to Dr. P. Cruz for her invaluable collaboration. Special thanks are also extended to Dr. de la Reza, and Prof. Ortolani for their productive scientific discussions. Particular acknowledge Prof. A. Díez Herrero (IGM-CSIC) for his encouragement this research endeavor. Additionally, thanks are extended to P. Mestanza for her technical support.

## Author Contributions

Félix Llorente de Andrés is the sole author. The author read and approved the final manuscript.

## Conflicts of Interest

The author declares no conflicts of interest.

## Appendix

*Table A1. List of GCs with their age values this work, Valcin´s work and Usher´s work.*

*Those GCs whose age calculated by VA is greater than 13.8Gyr are marked in bold*

| Cluster | [Fe/H] | This work age | VA age | Error VA | US age | Cluster | [Fe/H] | This work age | VA age | Error VA | US age |
|---|---|---|---|---|---|---|---|---|---|---|---|
| name | (--) | Gyr | Gyr | Gyr | Gyr | name | (--) | Gyr | Gyr | Gyr | Gyr |
| Arp 2 | -1.8 |  | 13.42 | +1.24-1.65 |  | NGC 6352 | -0.64 |  | 11.93 | +1.80-1.57 | 13 |
| IC4499 | -1.6 | 19.04 | 12.8 | +0.66-0.78 | 12 | NGC 6342 | -0.55 | 12.25 |  |  | 12.5 |
| Lynga7 | -0.62 |  | 10.82 | +2.12-1.54 | 4 | NGC 6356 | -0.4 | 13.44 |  |  | 12.75 |
| NGC 104 | -0.75 | 14.70-20.30 | 13.54 | +1.24 -1.65 | 12.75 | NGC 6362 | -0.99 | 8.99 | 13.58 | +0.82-0.61 | 12.5 |
| NGC 288 | -1.37 |  | 11.2 | 0.67 -0.67 | 12.5 | NGC 6366 | -0.6 |  | 12.15 | +1.46-1.46 |  |

| Cluster | [Fe/H] | This work age | VA age | Error VA | US age | Cluster | [Fe/H] | This work age | VA age | Error VA | US age |
|---|---|---|---|---|---|---|---|---|---|---|---|
| NGC 362 | -1.2 | 11.17 | 11.52 | +0.84-0.84 | 11.5 | NGC 6388 | -0.43 | 18.79 | 11.07 | +2.12-1.42 | 11.75 |
| NGC 1261 | -1.23 | 13.49 | 11.54 | +0.67-0.45 | 11.5 | NGC 6397 | -2.02 | 11.02 | 14.21 | +0.69-0.69 | 13.5 |
| NGC 1851 | -1.25 | 13.45 | 12.27 | +1.47--0.90 | 11 | NGC 6426 | -2.16 | | 13.92 | +0.96-1.12 | |
| NGC 2298 | -1.9 | | 13.89 | +0.88--0.63 | 13 | NGC 6441 | -0.37 | | 10.44 | +2.76-1.62 | 12 |
| NGC 1904 | -1.53 | 11.22 | | | 13 | NGC 6496 | -0.55 | | 10.86 | +2.11-1.64 | 12 |
| NGC 2808 | -1.19 | 10.48 | 10.93 | +1.2-1.08 | 11.5 | NGC 6535 | -1.95 | | 13.81 | +1.06-1.06 | 10.44 |
| NGC 3201 | -1.49 | 9.57 | 13.05 | +1.05-1.19 | 12 | NGC 6541 | -1.76 | | 13.51 | +0.86-0.65 | 13.25 |
| NGC 4147 | -1.66 | 13.50 | 13.02 | +0.50-0.54 | 12.75 | NGC 6522 | -1.34 | | | | 12.4 |
| NGC 4372 | | 11.30 | | | | NGC 6544 | -1.4 | 12.47 | | | 12.75 |
| NGC 4590 | -2.35 | 13.14 | 12.03 | -0.54+0.54 | 13 | NGC 6569 | -0.79 | 8.23 | | | 13 |
| NGC 4833 | -2.03 | 18.55 | 14.69 | +0.23-0.70 | 13 | NGC 6584 | -1.4 | 13.16 | 12.72 | +0.76-0.66 | 12.25 |
| NGC 5024 | -1.97 | | 13.31 | 0.66-0.57 | 13.25 | NGC 6624 | -0.69 | 13.31 | 11.29 | +1.90-1.27 | 13 |
| NGC 5053 | -2.45 | | 13.84 | +0.66-0.57 | | NGC 6637 | -0.69 | 13.21 | 12.85 | +1.35-1.35 | 12.5 |
| NGC 5139 | -1.6 | | 14.91 | -0.00+0.11 | 11 | NGC 6638 | -0.95 | 12.80 | | | 12.75 |
| NGC 5272 | -1.46 | | 12.6 | -0.66+0.66 | 12.5 | NGC 6642 | -1.26 | 13.51 | | | 11.5 |
| NGC 5286 | -1.73 | | 14.55 | -0.86+1.07 | 13 | NGC 6652 | -0.83 | 13.66 | 12.98 | +1.55-0.86 | 13.25 |
| NGC 5466 | -1.73 | | 12.31 | -0.60+0.40 | | NGC 6656 | -1.57 | | 14.54 | +0.36-0.97 | 13.5 |
| NGC 5634 | -1.87 | 15.79-21.59 | | | 13 | NGC 6681 | -1.52 | 9.60 | 13.87 | +0.73-0.83 | 13 |
| NGC 5694 | -1.74 | 12.34 | | | 13.6 | NGC 6712 | -0.94 | | | | 12 |
| NGC 5824 | -1.6 | 13.06 | | | 13 | NGC 6715 | -1.22 | | 12.22 | +1.9-1.43 | 13.25 |
| NGC 5946 | -1.22 | 13.28 | | | 12.75 | NGC 6717 | -1.15 | 13.362 | 11.65 | +1.5-1.71 | 13 |
| NGC 5904 | -1.28 | | 12.75 | +0.50-0.58 | 12.25 | NGC 6723 | -1.02 | 8.94 | 13.81 | +0.70-0.90 | 12.75 |
| NGC 5927 | -0.47 | | 8.33 | +1.98-1.13 | 12.25 | NGC 6752 | -1.58 | | 13.48 | +0.81-0.54 | 12.5 |
| NGC 5986 | -1.53 | 8.91 | 14.82 | +0.00-1.12 | 13.25 | NGC 6779 | -1.9 | | 14.85 | +0.08-0.76 | 12.75 |
| NGC 6093 | -1.78 | 12.31 | 13.83 | 0.96-0.72 | 13.5 | NGC 6809 | -1.93 | | 13.93 | +0.50-0.58 | 13.5 |
| NGC 6101 | -1.85 | | 13.22 | -0.66+0.66 | | NGC 6838 | -0.73 | 13.15 | 11.21 | +1.59-1.59 | 12.5 |
| NGC 6121 | -1.11 | | 13.01 | -1.01+1.01 | 12.5 | NGC 6864 | -1.1 | 12.41 | | | 11.25 |
| NGC 6144 | -1.73 | | 14.47 | -0.42+1.12 | 13.5 | NGC 6934 | -1.48 | 12.15 | | | 12 |
| NGC 6171 | -1.05 | 9.60 | | | 12.75 | NGC 6981 | -1.4 | 11.18 | 12.72 | +0.69-0.69 | 12.75 |
| NGC 6205 | -1.54 | 10.26 | 13.49 | +0.62+0.45 | | NGC 7006 | -1.55 | | 13.18 | +1.14-1.00 | 12.25 |
| NGC 6218 | -1.35 | 11.38 | 14.64 | +0.29-0.64 | 13.25 | NGC 7078 | -2.36 | 12.99 | 13.28 | +0.82-0.71 | 13.25 |
| NGC 6235 | -1.18 | 10.59 | | | 12.75 | NGC 7089 | -1.47 | | 13.08 | +0.85-0.85 | 12.5 |
| NGC 6254 | -1.55 | | 12.85 | -0.8+0.8 | 13 | NGC 7099 | -2.31 | 9.75 | 12.82 | +0.33-0.50 | 13.25 |
| NGC 6266 | -1.08 | 12.28 | | | 12.5 | palomar 1 | -0.7 | | 8.2 | +0.367-1.93 | |
| NGC 6273 | -1.53 | 18.51 | | | 13.2 | palomar 12 | -0.8 | | 9.94 | +0.92-0.73 | |
| NGC 6284 | -1.13 | | | | 12 | palomar 15 | -2 | | 13.97 | +0.88-1.76 | |
| NGC 6287 | -2.01 | 13.19 | | | 13.25 | pyxis | | | 14.84 | +0.00-3.28 | |
| NGC 6293 | -1.99 | 9.92 | | | 13 | rup106 | -1.48 | | 11.3 | +1.96-1.55 | 10.11 |

| Cluster | [Fe/H] | This work age | VA age | Error VA | US age | Cluster | [Fe/H] | This work age | VA age | Error VA | US age |
|---|---|---|---|---|---|---|---|---|---|---|---|
| NGC 6304 | -0.51 | 13.27 | 8.67 | +1.80-1.80 | 12.75 | terzan7 | -0.6 | | 8.1 | +1.96-1.40 | |
| NGC 6341 | -2.33 | | 13.3 | +0.6-0.6 | | terzan8 | -2.255 | | 13.48 | +0.90-0.77 | |